# A Numerical Study of Lid Driven Cavity with Mixed Convection

By:
 Aditya Shankar Garg
 Ishan Singh
 (Indian Institute of Technology, Delhi)

**Abstract**

Direct Numerical Simulation have been carried out for a two dimensional flow in a Lid driven cavity at Reynolds number 5000 and Prandtl number 7 with water as the working fluid. Both the side walls of the enclosure are insulated(i.e. adiabatic boundary condition), while the bottom plate is at higher temperature and the top wall is at colder temperature. Effects of heating of the bottom wall and movement of the top lid have been investigated by conducting numerical simulations at different Richardson numbers by varying from low and moderate magnitudes within the limits of Boussinesq-approximation. Three standard cases has been compared, in the first case heating effects are not taken into account and only the flow due to shear action of the plate is studied. In the second case only the heating effects are taken into account and shear effects are neglected. In the third case effects of both heating and shear action is taken into consideration(i.e. mixed convection). Drag force on the moving plate is calculated in all the three cases and effect of temperature on the drag force is studied. For running the above simulation a code has been developed which is validated by comparing the results with Ghia et al for non-heating case.

**Introduction**

Some important definitions of non-dimensional numbers

Richardson number is the ratio of buoyancy term to that of inertia term. It is mostly used in the context of mixed convection. If the Richardson number is much less than unity, buoyancy is unimportant in the flow. If it is much greater than unity, buoyancy is dominant. If Richardson number is of order unity then the flow is likely to be buoyancy-driven.

$$\text{Richardson Number } (Ri) = \frac{\text{Bouyancy Force}}{\text{Inertia Force}} = \frac{Gr}{Re^2}$$

**Grashof number** approximates the ratio of the buoyancy to viscous force acting on a fluid. It frequently arises in the study of situations involving natural convection and is analogous to Reynolds number.

$$\text{Grashof Number}(Gr) = \frac{\text{Bouyancy Force}}{\text{Viscous Force}} = \frac{g\beta l \Delta T_o}{U_o^2} \left(\frac{U_o l}{\nu}\right)^2$$

**Rayleigh number** is defined as the product of the Grashof number, which describes the relationship between buoyancy and viscosity within a fluid, and the Prandtl number, which describes the relationship between momentum diffusivity and thermal diffusivity.

Rayleigh Number$(Ra) = Gr \cdot Pr$

If $Ra < Ra_{critical}$ : conduction heat transfer   and   if $Ra > Ra_{critical}$ : convection heat transfer is dominant.

**Nusselt number** is the ratio of convective to conductive heat transfer across (normal to) the boundary. In the context convection includes both advection and diffusion.

$$\text{Nusselt Number}(Nu) = \frac{\text{convective heat transfer}}{\text{conductive heat transfer}} = \frac{hl}{k}$$

For forced convection, Nu is independent of Grashof number → $Nu = f(Re, Pr)$

For natural convection, Nu is independent of Reynolds number → $Nu = f(Gr, Pr)$

**Prandtl number** is the ratio of momentum diffusivity to thermal diffusivity.

$$\text{Prandtl Number} = \frac{\text{viscous diffusion rate}}{\text{thermal diffusion rate}} = \frac{\nu}{\alpha} = \frac{\mu C_p}{k}$$

*For air, Pr = 0.71*

*For water, Pr = 7.56*

Small values of the Prandtl number, Pr << 1 means the thermal diffusivity dominates. Whereas large values of Prandtl number signifies that the momentum diffusion dominates.

The Prandtl number of gases are about 1, which indicates that both momentum and heat dissipate through the fluid at about the same rate. Heat diffuses very quickly in liquid metals (Pr << 1) and very slowly in oils ( Pr >> 1) relative to momentum. Consequently thermal boundary layer is much thicker for liquid metals and much thinner for oils relative to viscous boundary layer.

**Literature Survey for the Project**

**Bejan et al** (1992) performed numerical simulation to calculate entropy generation through heat and fluid flow.

**Jang et al.** (1992) investigated Prandtl Number effects on laminar mixed convection heat transfer in a lid-driven cavity. The flow and heat transfer is calculated in a bottom heated lid-driven square cavity flow.The effects of Prandtl number on the flow structure and the and the heat transfer in the cavity are studied for laminar ranges of Reynolds number and Grashoff number. They found that the influence of buoyancy on the flow and heat transfer in the cavity is found to be more pronounced for higher value Pr if Re and Gr are kept constant. The natural convection effects are always assisting forced convection heat transfer and the extent of contribution is a function of Pr and Ri.

**Hyun et al.** (1992) did the numerical simulation of mixed convection in a driven cavity with a stable vertical temperature gradient. A stabilising externally imposed vertical temperature differential $(T_{top} - T_{bottom}) > 0$ imposed is enforced upon the system boundary. For Ri << 1 flow characteristics are similar to that of the conventional lid-driven cavity of non-stratified fluid .The isotherms are clustered only close to the top and bottom walls. In the bulk of the central region of the cavity, fluids are well mixed and temperature variation is small. For Ri>>1 much of the middle and bottom portion of the fluid is stagnant. In this near stagnant region isotherm are fairly horizontal and vertically linear temperature distributions prevail. The only zone close to the top sliding wall fluids is well mixed and zone of fairly uniform temperature is noticeable. When Pr is low, the temperature field in the interior tends to a vertically linear profile throughout the entire cavity. When Pr is high, the temperature profile is uniform in the upper part of the cavity. In the bottom part temperature distribution shows a vertically linear profile.

**Aydin et al.** (1999) studied aiding and opposing mechanisms of mixed convection in shear and buoyancy-driven cavity. The focus was on the interaction of the forced convection induced by the moving wall with the natural convection induced by the buoyancy. Two orientation of thermal boundary conditions at the cavity walls are considered in order to simulate the aiding and opposing buoyancy mechanism. Results were obtained for different values of mixed convection parameter, Ri, in the range .01-100 at Re=100. With the increasing value of Ri three different heat transport regimes were defined as forced convection, mixed convection, natural convection. The mixed convection range for Ri for the opposing buoyancy case was seen to wider than that of the aiding buoyancy case.

**Oztop et al.** (2004) investigated mixed convection in a two-sided lid-driven differentially heated square cavity. The different cases have been performed by the direction of the movement of the vertical walls. It is found that both Ri and direction of moving walls affect the fluid flow and heat transfer in the cavity. For Ri<1 the influence of moving walls on the heat transfer is the same when the move in opposite direction regardless of which side moves upward. For the cases of opposing buoyancy and shear forces and for Ri>1 the heat transfer is somewhat better due to formation of secondary cells on the walls and a counter-rotating cell at the centre.

**Kieft et al.** (2007) studied vortex shedding mechanism due to heat input near wake. A brief study has also been conducted for vorticity due to temperature gradient.

**Cheng et al.** (2010) studied the effect of temperature gradient orientation on the characteristics of mixed convection flow in a lid-driven cavity. It is found that both Richardson Number and direction of temperature gradient affect the flow patterns, heat transport process and heat transfer rates in the cavity. Computed average Nusselt Number indicates that the heat transfer rate increases with decreasing Ri regardless the orientation of temperature gradient imposed.

**Cheng et al.** (2010) investigated the characteristics of mixed convection heat transfer in a lid driven square cavity with various Ri and Pr. To what extent the mixed convection flows would change from laminar to chaos has been investigated with the help of time traces of total kinetic energy and average nusselt number. Hopf bifurcation occurs at Ri =1 and 10 at certain Reynolds number which is a function of fluid Prandtl number. The average Nusselt Numbers are reported to illustrate the influence of flow parameter variation on heat transfer.

**Teamah, Sorour, El-Maghlany, Amr Afifi et al** (2013) conducted direct numerical simulations on an inclined lid driven cavity flow by considering a shallow cavity.

**Khorasanizadeh et al** (2014) studied the entropy generated due to heating effect and entropy generated due to viscous effect in mixed convection.

**Mohammad et al and Viskanta et al.** did the numerical simulations for the transient state problem in which the upper lid is given a sinusoidal motion.

**Problem Definition** :

In this project we consider only two dimensional setup and also to incorporate a symmetrical set up we take the shape of the cavity to be a square. To ignore the third dimension we say that it is very long. The geometry of the cavity is as follows :

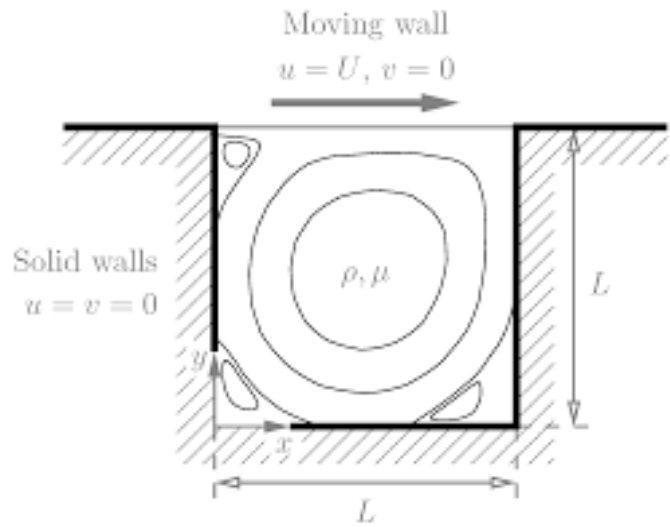

The top wall moves along the positive x-direction with a constant velocity of $U_o$ and all the other walls are stationary. The walls are at some known reference temperature and the cavity is filled with water. The thermo-physical properties of the working fluid is assumed to be constant except the mass density which will vary according to the Boussinesq approximation. The fluid is taken to be newtonian $\left(\tau = -\mu \frac{\partial v}{\partial x}\right)$ and incompressible $(\nabla . v = 0)$.

**Mathematical formulation :**

Continuity equation :

$$\frac{\partial u}{\partial x} + \frac{\partial v}{\partial y} = 0$$

X-direction momentum equation :

$$\frac{\partial u}{\partial t} + u\frac{\partial u}{\partial x} + v\frac{\partial u}{\partial y} = \frac{-1}{\rho_o}\frac{\partial p}{\partial x} + \nu\left(\frac{\partial^2 u}{\partial x^2} + \frac{\partial^2 u}{\partial y^2}\right)$$

Y-direction momentum equation :

$$\frac{\partial v}{\partial t} + u\frac{\partial v}{\partial x} + v\frac{\partial v}{\partial y} = \frac{-1}{\rho_o}\frac{\partial p}{\partial y} + \nu\left(\frac{\partial^2 v}{\partial x^2} + \frac{\partial^2 v}{\partial y^2}\right) + g\beta(T - T_c)$$

Energy Equation for the system :

$$\frac{\partial T}{\partial t} + u\frac{\partial T}{\partial x} + v\frac{\partial T}{\partial y} = \alpha\left(\frac{\partial^2 T}{\partial x^2} + \frac{\partial^2 T}{\partial y^2}\right)$$

Non - dimensionalize the given equations using the following substitutions :

$$X = \frac{x}{L},\ Y = \frac{y}{L},\ U = \frac{u}{U_o},\ V = \frac{v}{U_o},\ \theta = \frac{T - T_c}{T_h - T_c},\ \tau = \frac{tU_o}{L},\ P = \frac{p}{\rho U_o^2}$$

**The governing equations for such a set up is**

Continuity :

$$\frac{\partial U}{\partial X} + \frac{\partial V}{\partial Y} = 0$$

X - direction :

$$\frac{\partial U}{\partial \tau} + U\frac{\partial U}{\partial X} + V\frac{\partial U}{\partial Y} = -\frac{\partial P}{\partial X} + \frac{1}{Re}\left(\frac{\partial^2 U}{\partial X^2} + \frac{\partial^2 U}{\partial Y^2}\right)$$

Y - direction :

$$\frac{\partial V}{\partial \tau} + U\frac{\partial V}{\partial X} + V\frac{\partial V}{\partial Y} = -\frac{\partial P}{\partial Y} + \frac{1}{Re}\left(\frac{\partial^2 V}{\partial X^2} + \frac{\partial^2 V}{\partial Y^2}\right) + Ri\theta$$

Energy Equation :

$$\frac{\partial \theta}{\partial \tau} + U \frac{\partial \theta}{\partial X} + V \frac{\partial \theta}{\partial Y} = \frac{1}{Re \cdot Pr} \left( \frac{\partial^2 \theta}{\partial X^2} + \frac{\partial^2 \theta}{\partial Y^2} \right)$$

**General Computational Formulation of the Problem**

The governing equations for the lid driven cavity's thermally driven flow represent a system of non-linear and coupled partial differential equations that are elliptical in nature. To conduct direct numerical simulations we used a large number of space and time scales and hence we have a fine grid near the walls as well as smaller time-steps. Our idea was to modify the Simplified Marker and Cell algorithm to incorporate the pressure correction.

Temporal Discretisation is done using the Euler Scheme that can be written as the following equation :

$$\frac{u^{(n+1)} - u^{(n)}}{\Delta t} = \frac{-1}{\rho} \nabla p^{(n+1)} - u^{(n)} \cdot \nabla u^{(n)} + \nu \nabla^2 u^{(n)}$$

The simulation is started with n = 0 and the initial condition used to populate the velocity field $u^{(0)}$. Then the equation is subsequently used while using $u \cdot \Delta t / \Delta x < 1$ also known as the CFL condition.

It has to be noted that the above equation doesn't incorporate the continuity equation and hence divergence is not necessarily vanishing. To take care of this we use a two step algorithm as a modification to the SMAC, also known as the prediction-corrector step.

In the first step we have to compute the intermediate velocity by solving the momentum equation but not incorporating the effect of pressure :

$$\frac{u^* - u^{(n)}}{\Delta t} = -u^{(n)} \cdot \nabla u^{(n)} + \nu \nabla^2 u^{(n)}$$

In the second step also known as the corrector step we obtain the new velocity and include the influence of pressure :

$$\frac{u^{(n+1)} - u^*}{\Delta t} = \frac{-1}{\rho} \nabla p^{(n+1)}$$

U-momentum discretisation is done using finite difference method which approximate the derivatives using the neighbouring values. The idea is to discrete the predictor step equation given by :

$$u^* = u^{(n)} - \Delta t \left( \nu \left( \frac{\partial^2 u^{(n)}}{\partial x^2} + \frac{\partial^2 u^{(n)}}{\partial y^2} \right) - \left( u^{(n)} \frac{\partial u^{(n)}}{\partial x} + v^{(n)} \frac{\partial u^{(n)}}{\partial y} \right) \right)$$

In the above equation the viscous and the convective terms are discretised for i, j cell using :

$$\frac{\partial^2 u}{\partial x^2} = \frac{u(i-1,j) - 2u(i,j) + u(i+1,j)}{\Delta x^2}$$

$$\frac{\partial^2 u}{\partial y^2} = \frac{u(i,j-1) - 2u(i,j) + u(i,j+1)}{\Delta y^2}$$

$$u \frac{\partial u}{\partial x} = u(i,j) \left( \frac{u(i+1,j) - u(i-1,j)}{2\Delta x} \right)$$

$$v \frac{\partial u}{\partial y} = \frac{1}{4}(v(i-1,j) + v(i,j) + v(i-1,j+1) + v(i,j+1)) \left( \frac{u(i,j+1) - u(i,j-1)}{2\Delta y} \right)$$

A similar scheme can be used for the other component of velocity $v$ by replacing $u$ by $v$ and interchanging $x$ by $y$.

**Results and Discussion** :

**Horizontal Velocity Profile at Mid Section of Cavity at Various Richardson Numbers :**

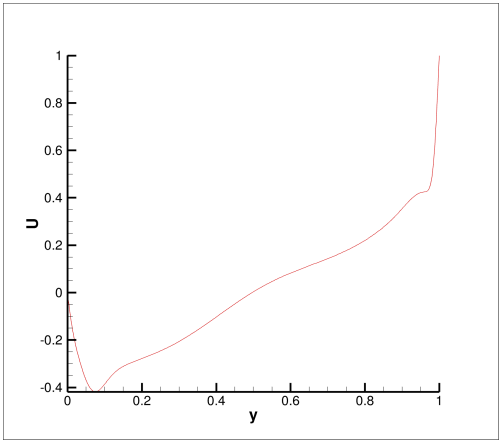

Fig 1 : U vs y for Ri = 0, Re = 5000

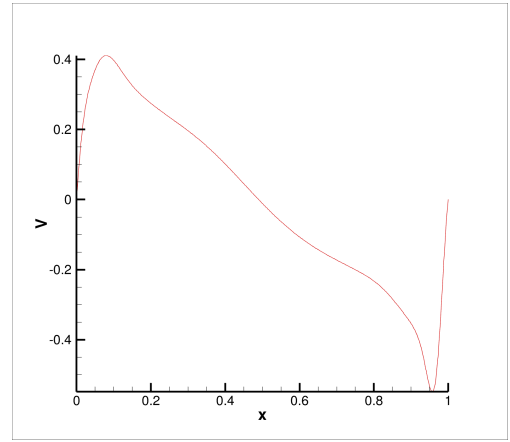

Fig 2 : V vs x for Ri = 0, Re = 5000

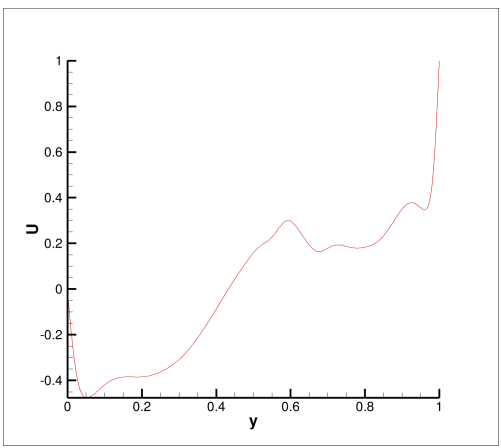

Fig 3 : U vs y for Ri = 1, Re = 5000

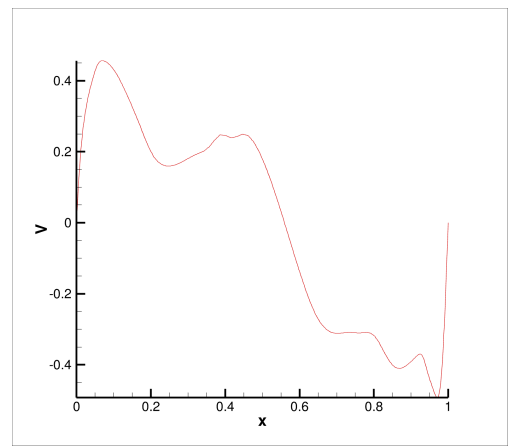

Fig 4 : V vs x for Ri = 1, Re = 5000

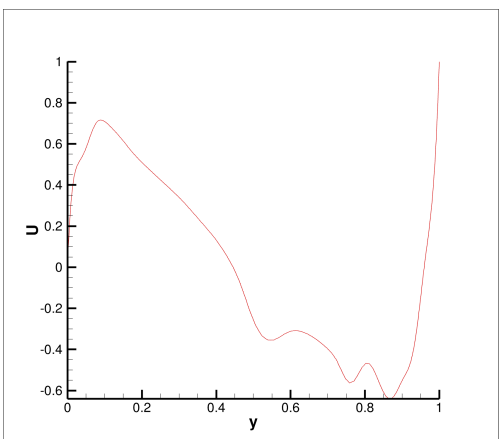

Fig 5 : U vs y for Ri = 5, Re = 5000

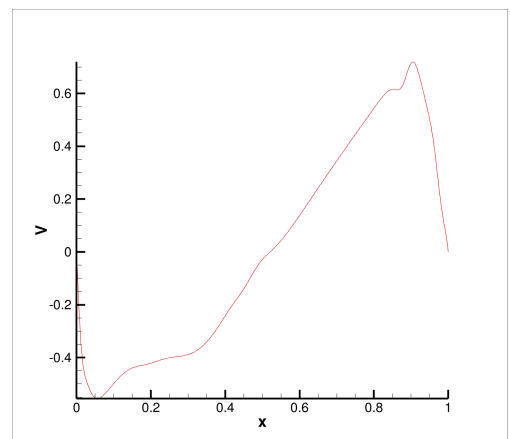

Fig 6 : V vs x for Ri = 5, Re = 5000

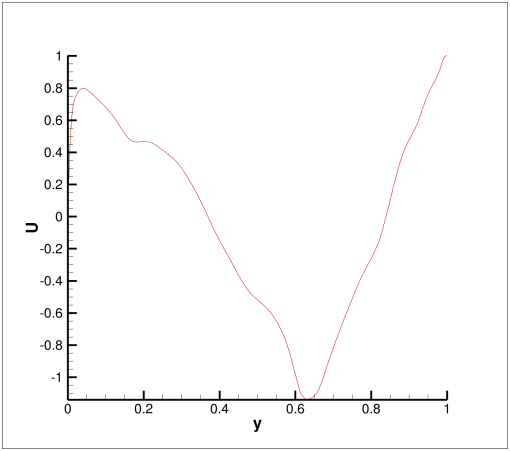

Fig 7 : U vs y for Ri = 10, Re = 5000

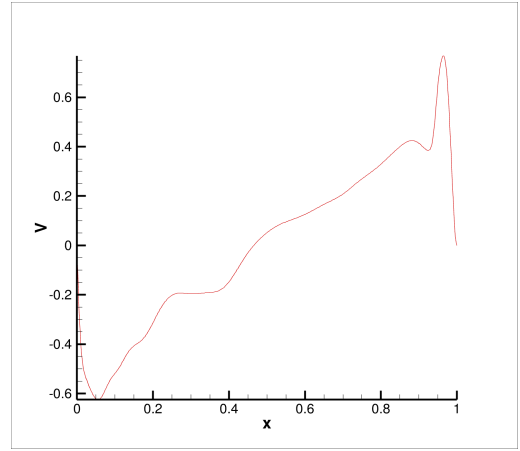

Fig 8 : V vs x for Ri = 10, Re = 5000

**Temperature Profile for the Cavity for various Richardson Numbers :**

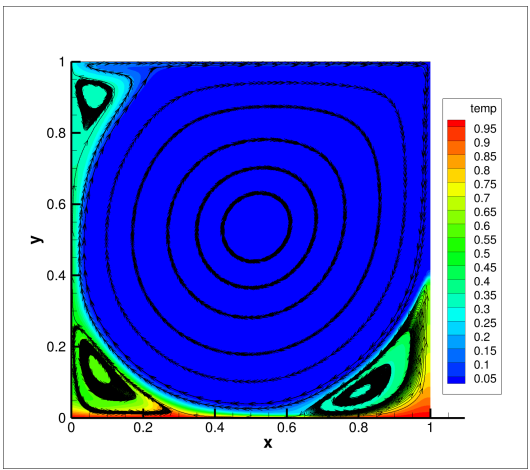

(a)

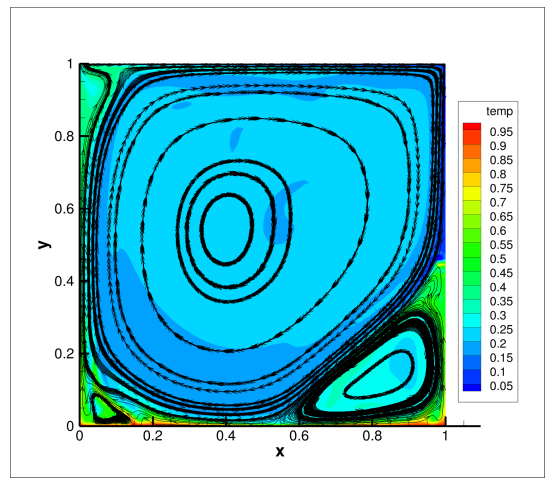

(b)

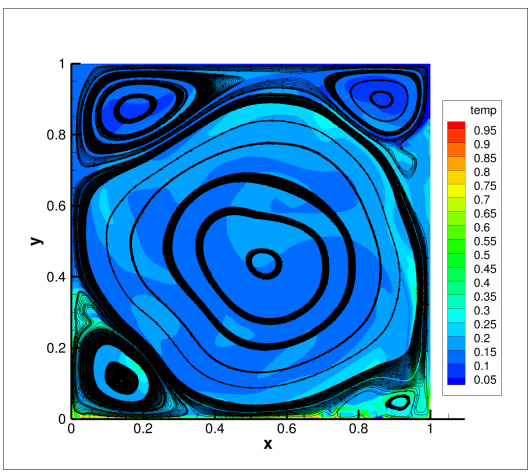

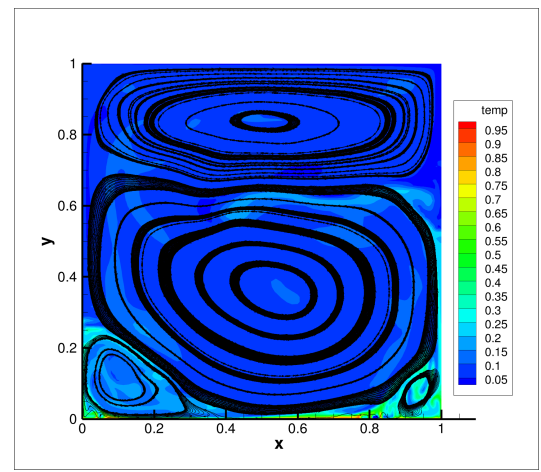

Fig 9 : Temperature Contours for (a) Ri = 0, (b) Ri = 1, (c) Ri = 5, (d) Ri = 10

The velocity profiles at the mid-section were matched along with Ghia et al.'s results (1983) for the lid driven cavity (non heating case) and found to be valid.

From the temperature contours we can see that at low values of Richardson number (Ri=0) the temperature is almost constant in the major part of the cavity with a dominating clockwise vortex/ circulation which indicates that inertia is dominant at Ri=0, there are three minor vortices generated (two at the bottom and one at top right corner). At Ri=1 there is one primary vortex in the centre while the strength of the bottom-right vortex increases and bifurcation is achieved. Bifurcations are clearly visible at the Ri=5, 10 cases. The primary circulation in these cases becomes counter-clockwise due to the fact that buoyancy force starts to dominate the inertia flow and there is deficit of kinetic energy needed to homogenise the fluid.

**Vorticity Profiles for the Lid Driven Cavity at Various Richardson Numbers :**

Vorticity is a pseudo-vector field that describes the local spinning motion of a continuum near some point. More precisely, the vorticity is a pseudo-vector field $\omega$, defined as the curl of the flow velocity vector.

$$\vec{\omega} = \nabla \times \vec{V}$$

Vorticity is a useful tool to understand how the ideal potential flows solutions can be perturbed to model real flows.

The vorticity contours are plotted for all the Richardson numbers discussed above and are shown below. We can see the fluid flow bifurcations clearly for Ri=1, Ri=5 and Ri=10. Increasing Ri increases the strength of vortex.

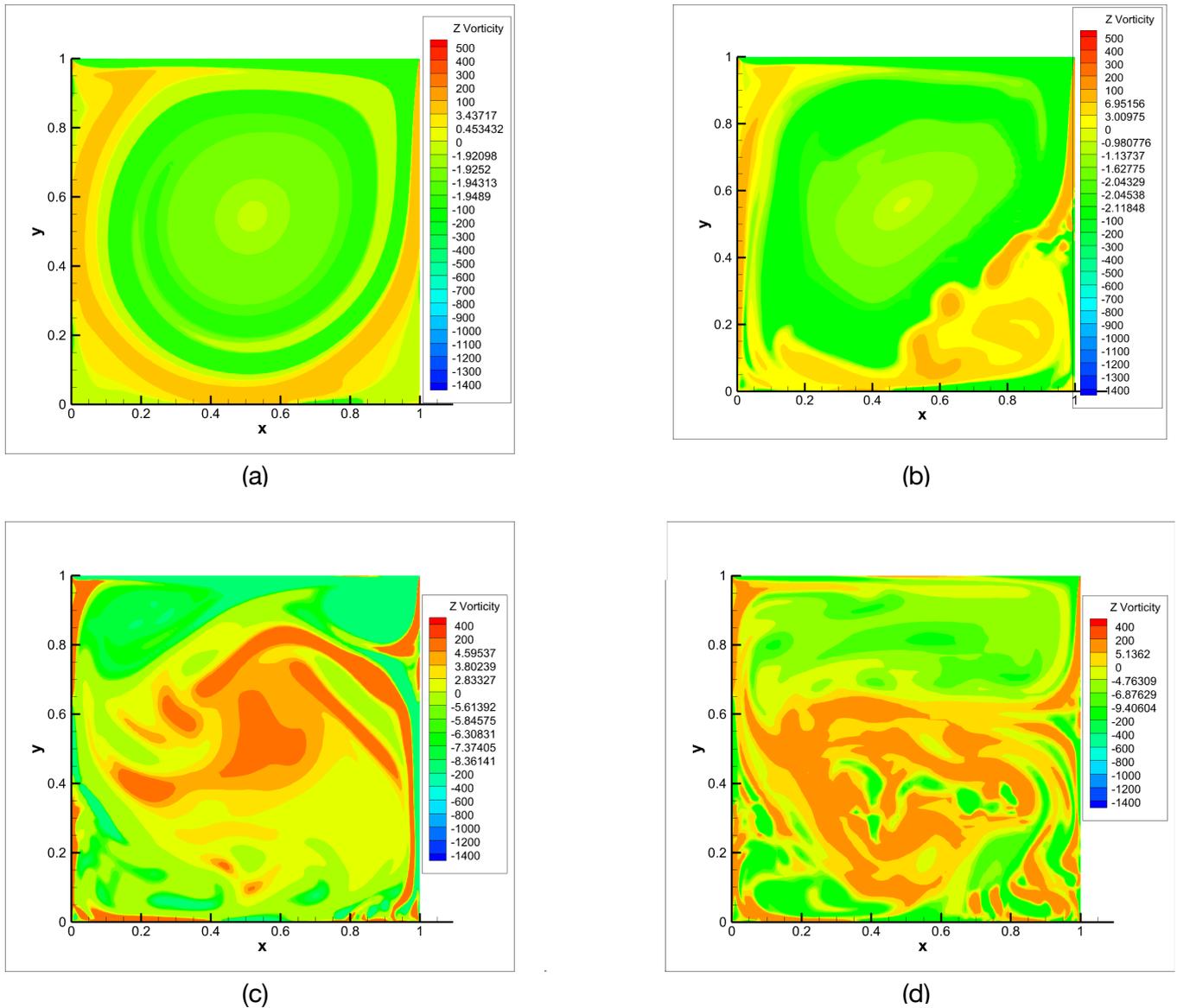

Fig 10 : Vorticity Contours for (a) Ri = 0, (b) Ri = 1, (c) Ri = 5, (d) Ri = 10

**Nusselt Number variations for the bottom wall with varying Richardson Number :**

It is the ratio of convective to conductive heat transfer across the boundary. In this context convection includes both advection and diffusion. The conductive component is measured under the same conditions as the heat convection but with a (supposed) stagnant fluid. A Nusselt number close to one, conduction and convection are equal contributors and depicts laminar flow.

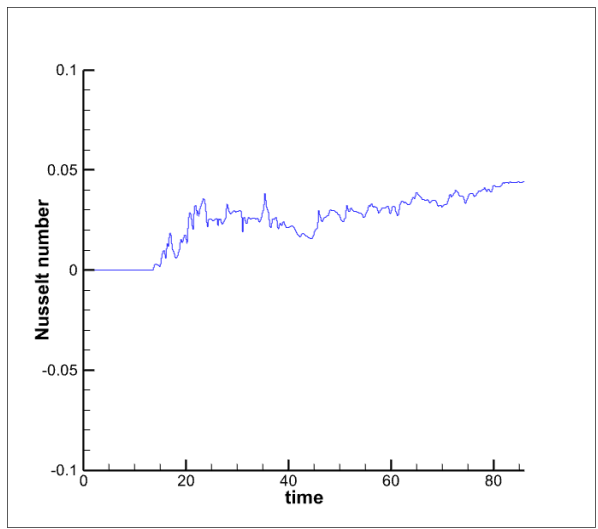

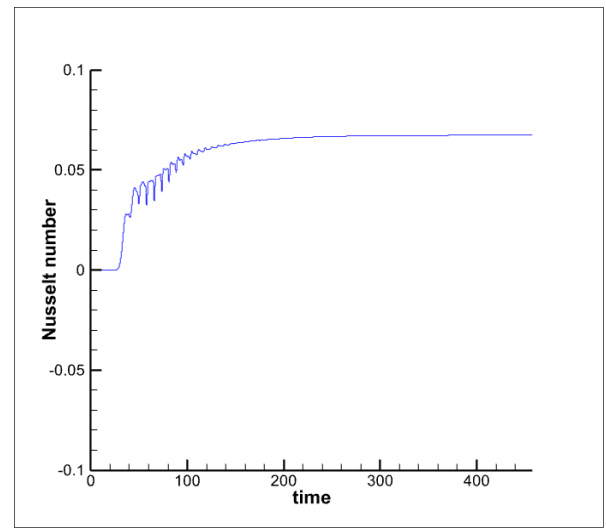

(a) (b)

Fig 11 : The time history of nusselt number for the bottom wall for cases (a) Ri = 0, (b) Ri =5

**Wall Shear Stress Simulations :**

A fluid at rest cannot resist shearing forces. Under the action of such forces it deforms continuously, however small they are. The resistance to the action of shearing forces in a fluid appears only when the fluid is in motion. This implies the principal difference between fluids and solids. For fluids the shear stress τ is a function of the rate of strain. For Newtonian fluids, shear stress is directly proportional to the rate of strain. Mathematically,

$$\tau = \frac{1}{Re} \frac{\partial U}{\partial Y}$$

Simulations for the shear stress was done for 4 different Ri at distinct time steps and we will plot the variation of shear stress for the range $0.2 \leq X \leq 0.8$ at the final time step:

Ri = 0 :

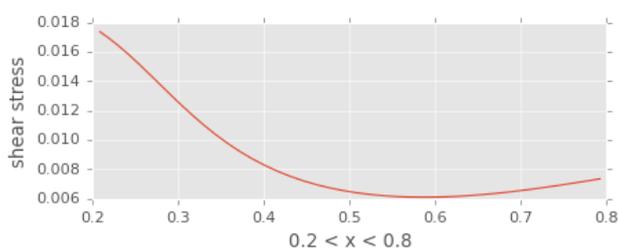

Ri = 1:

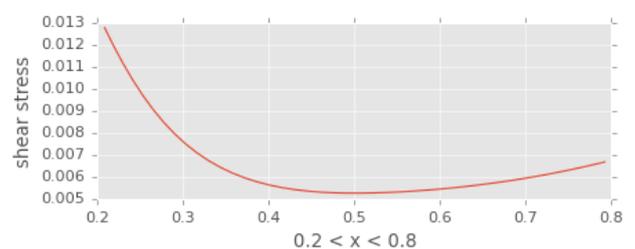

Ri = 5:

Ri = 10:

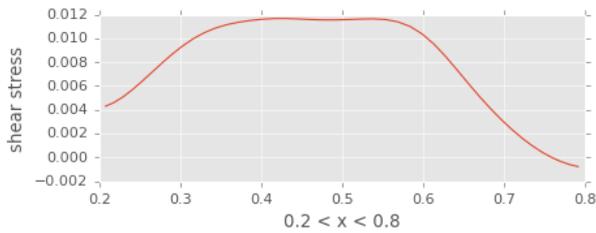 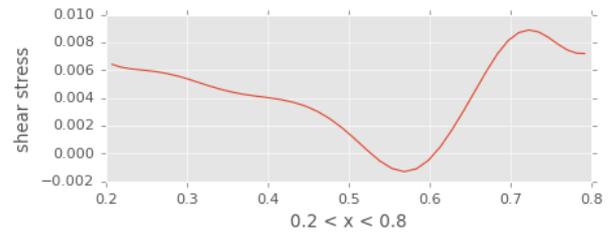

For Ri = 5 the shear stress becomes negative close to x=0.7 and x=0.8 as the vortex formed at these points close to the top wall is anticlockwise in direction. At the left corner the shear stress is maximum because there is sudden change in velocity i.e. from zero to velocity of the top moving wall. So, there are strong velocity gradient present at the left corner of the cavity. Also, the shear stress is maximum at the top right corner(i.e. x=1) because the velocity suddenly reduces to zero i.e. stagnation point. In the intermediate portion of the cavity the shear stress is close to zero due to the very low value of the dynamic viscosity of water i.e. $8.9 \times 10^{-4} Pa - s$. For Ri = 0,1 we observe that the shear stress on the moving lid is always positive in this range. For Ri = 10 we observe that shear stress attains a local maxima and minima in the range.

These shear stress distributions at different time steps are used to calculate the drag force on the top moving wall. Drag force is nothing but the area under the shear stress curve :

$$F_d = \int_0^\infty \tau dx$$

The integral was calculated using the python solver for the trapezoidal scheme the results are as follows :

The values of non-dimensional drag for different Ri at different time steps

| Snapshot Number | $F_d$ (Ri = 0) | $F_d$ (Ri = 1) | $F_d$ (Ri = 5) | $F_d$ (Ri = 10) |
|---|---|---|---|---|
| 10 | 0.0149 | 0.0133 | 0.0129 | 0.0158 |
| 20 | 0.0149 | 0.0132 | 0.0140 | 0.0113 |
| 40 | 0.0149 | 0.0144 | 0.0132 | 0.0113 |
| 80 | 0.0149 | 0.0134 | 0.0130 | 0.0113 |

It can be seen from the simulation results that for Ri = 0 the drag force on the moving lid remains more or less the same with time, for Ri = 1 the drag force first decreases with time and then increases till reaching a maxima and then again decreases asymptotically, for Ri = 5 the drag force first increases reaches a maximum and then decreases with time asymptotically, for Ri = 10 the drag force decreases with increasing time and then flatlines to the value of about 0.0113799.

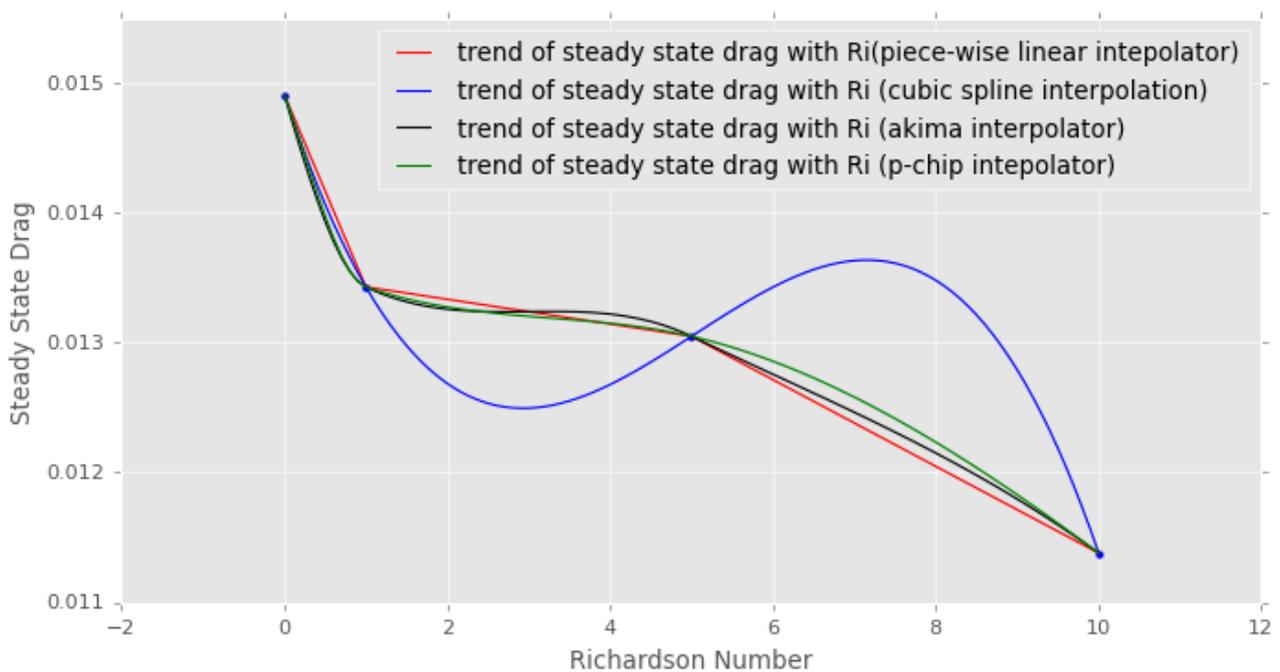

Fig 12 : Variation of steady state drag force with Richardson Number for the Lid Driven Cavity

**Conclusions :**

From this study we concluded that for lid driven cavity flows of $Re = 5000$, $Ri = 1, 5, 10$ we expect fluid bifurcation (which means that the major vortex will be split up). This can be explained using the fact that buoyancy force starts to dominate the inertia flow and there is deficit of kinetic energy needed to homogenise the fluid.

The time history of average nusselt number for the bottom plate indicates that the nusselt number attains a stable value as we progress through time.

We also can used a simple first order approximation for the partial y-derivative of velocity to calculate the shear stress distribution. From the shear stress distribution we can conclude that for flows with no major bifurcation the shear stress on the plate is never negative and as bifurcation become more and more prominent the stress starts becoming negative in nature.

On integrating this we get the drag force. We observed that the value of the drag force is reducing as Richardson number increases this can be explained by the fact that we provide higher amount of thermal energy to the fluid which is able to reduce the drag force that the lid experiences. This shows that the drag force reduction is mainly due to effect of extensive heating on the viscous properties of working fluid.

| Richardson number | Thermal Energy | Frictional Work | % effectiveness |
|---:|---:|---:|---:|
| 0 | 0 | 0.0149 | N/A |
| 1 | 1 | 0.0134 | 0.036 |
| 5 | 5 | 0.0130 | 0.072 |
| 10 | 10 | 0.0113 | 0.36 |